\documentclass[%
 reprint,
 amsmath,amssymb,
aps,
]{revtex4-2}

\usepackage{graphicx}
\usepackage{dcolumn}
\usepackage{bm}
\usepackage{color}

\begin{document}

\preprint{APS/123-QED}

\title{Linear superposition in the general heavenly equation}

\author{S. Y. Lou}
\affiliation{%
	School of Physical Science and Technology, Ningbo University, Ningbo, 315211, China
}%
\author{Xiazhi Hao}%
\email{Corresponding author. haoxiazhi@zjut.edu.cn}
\affiliation{%
	College of Science, Zhejiang University of Technology, Hangzhou, 310014, China
}%




\date{\today}

\begin{abstract}
Evidently, the linear superposition principle can not be exactly established as a general principle in the presence of nonlinearity, and, at the first glance, there is no expectation for it to hold even approximately. In this letter, it is shown that the balance of different nonlinear effects describes what linear superpositions may occur in nonlinear systems. The heavenly equations are of significance in several scientific fields, especially in relativity, gravity, field theory, and fluid dynamics. A special type of implicit shock wave solution with three two-dimensional arbitrary functions of the general heavenly equation is revealed. Restrict the two-dimensional arbitrary functions to some types of one-dimensional arbitrary functions, it is found that the nonlinear effects can be balanced such that the ``impossible" linear superposition solutions can be nontrivially constituted to new solutions of the general heavenly equation.
\end{abstract}

\maketitle


The richness of nonlinear phenomena allows us to discover unexpected
relations among various mathematical and physical problems. For instance, in Ref. \cite{AA}, Alexandrov provides a simple and surprising relationship between two classical integrable systems, the Korteweg-de Vries (KdV) and type B Kadomtsev–Petviashvili (BKP) hierarchies. In Ref. \cite{Hao}, we find that there are at least five integrable hierarchies including three types of the KdV hierarchies, the special Svinolupov-Sokolov hierarchy, the Sharma-Tasso-Olver hierarchy and the Sawada-Kortera hierarchy solve the BKP hierarchy. In this letter, we show another unexpected phenomena, the linear superposition principle for nonlinear systems \cite{LS1,LS6,LS2,LS5,LS7,LS4,LS3,LS}. To the best of our knowledge, this phenomena is rarely found in nontrivial nonlinear physical systems.

The principle of linear superposition is the cornerstone of linear theory and is probably the main reason why it has developed so well and applied to many areas of science.
While linear behavior is an exception in Nature, nonlinearity is generic. What makes the analysis of nonlinear dynamics challenging is a well-known result that the principle of linear superposition does not generally apply to nonlinear systems \cite{nls1,nls2,nls3,mssp}. The lack of linear superposition principle prevents the construction of new solutions and it is generally impossible to linearly combine known solutions into new ones. It is known that for nonlinear systems, stable localized excitations can be formed from the balance of two or more physical effects such as the dispersion and the self-steepening effects \cite{prs,prs1,prs2,prs3}.
Inspired by the principle of balance, one may then ask whether a linear superposition of special solutions can under appropriate circumstances result in another solution of the same nonlinear system. To check this, we are trying in finding some types of special solutions such that the ``impossible" linear superposition theorem for nonlinear systems becomes possible provided that nonlinear effects are balanced, and this constitutes the subject of this letter.
The idea will be elucidated by investigation of the general heavenly equation (GHE) though could be exploited to treat other physical models.

The GHE was introduced as a result of classification of integrable symplectic Monge-Amp\`ere equations in four dimensions \cite{HE1,HE5}. It is one in the list of six heavenly type equations, and it is remarkably simple and symmetric, having the form
\begin{equation}
au_{xt}u_{yz}+bu_{yt}u_{xz}+cu_{zt}u_{xy}=0,~a+b+c=0,\label{GHEu}
\end{equation}
where $a,b$ and $c$ are arbitrary constants satisfying one linear relation given above, subscripts denote partial derivatives. $u=u(x,y,z,t)$ is a holomorphic function of four complex variables.
The GHE \eqref{GHEu} may be further cast in terms of new variables
\begin{eqnarray}
p=u_x,\ q=u_y,\ r=u_z\label{pqr}
\end{eqnarray}
as
\begin{eqnarray}
&&a\{r,\ p\}_{yt}+b\{r,\ q\}_{xt}=0,\label{GHE}\\
&&p_y=q_x,\label{GHEpy}\\
&&p_z=r_x\label{GHEpz}
\end{eqnarray}
with the help of the usual Poisson bracket $\{A,\ B\}_{\alpha\beta}$
being defined by
\begin{eqnarray}
\{A,\ B\}_{\alpha\beta}\equiv A_{\alpha} B_{\beta}-A_{\beta}B_{\alpha}. \label{PB}
\end{eqnarray}
The GHE \eqref{GHE} may be regarded as an invariant form of the self-dual Einstein field equations since any of the known heavenly equations such as the Husain's equation \cite{HE2}, the first and second Pleba\'nski's equations \cite{HE3} governing self-dual Einstein spaces can be mapped to the GHE \cite{HE}.  These so-called heavenly equations make up an important class of multi-dimensional integrable systems since they are obtained by a reduction of the Einstein equations with Euclidean (and neutral) signature for (anti-) self-dual gravity which includes the theory of gravitational instantons \cite{HE4}. The GHE \eqref{GHE} is an important example of such equations.
The main reason that allows for the application of superposition principle to GHE \eqref{GHE} is the balance of the nonlinear cross terms. This is possible only for certain types of solutions that have this property.

Let $\{p_1,\ q_1,\ r_1\}$ and $\{p_2,\ q_2,\ r_2\}$ being two solutions of GHE \eqref{GHE},
we wish to examine if and when a linear superposition
\begin{equation}
p=a_1p_1+a_2p_2,\ q=a_1q_1+a_2q_2,\ r=a_1r_1+a_2r_2,\label{LS}
\end{equation}
where $a_1$ and $a_2$ are arbitrary constants, could exactly solve GHE \eqref{GHE}. To establish this superposition, all we need to find is the condition under which the cross terms  in the nonlinear terms vanish. A description of the result is given in the following statement.\\
\bf  Theorem. \rm \it	If $\{p_1,\ q_1,\ r_1\}$ and $\{p_2,\ q_2,\ r_2\}$ satisfy GHE \eqref{GHE}--\eqref{GHEpz} with a further condition
\begin{equation}
a\{r_2,\ p_1\}_{yt}
+a\{r_1,\ p_2\}_{yt}
+b\{r_2,\ q_1\}_{xt}
+b\{r_1,\ q_2\}_{xt}=0,\label{GHE12}
\end{equation}
then the linear superposition \eqref{LS}
is also a solution of GHE \eqref{GHE}--\eqref{GHEpz}.\rm\\
\bf \em Proof. \rm The correctness for the equations \eqref{GHEpy} and \eqref{GHEpz} are trivial because of their linearity. Substituting the linear superposition \eqref{LS} into \eqref{GHE} yields
\begin{eqnarray}
&&a_1^2 \big[a\{r_1,\ p_1\}_{yt}+b\{r_1,\ q_1\}_{xt}\big]\nonumber\\
&&
+a_2^2 \big[a\{r_2,\ p_2\}_{yt}+b\{r_2,\ q_2\}_{xt}\big]\nonumber\\
&&
+a_1 a_2 \big[a \{r_2,\ p_1\}_{yt}
+a \{r_1,\ p_2\}_{yt}\nonumber\\
&&
+b\{r_2,\ q_1\}_{xt}
+b \{r_1,\ q_2\}_{xt}\big]=0.\label{GHE13}
\end{eqnarray}
It is clear that the first two terms of \eqref{GHE13} vanish because  $\{p_1,\ q_1,\ r_1\}$ and $\{p_2,\ q_2,\ r_2\}$ are solutions of \eqref{GHE}.
The remaining term coincides with the identity \eqref{GHE12}.
The theorem is proven.

The theorem above makes possible the linear superposition between two solutions of the GHE resulting from balance condition \eqref{GHE12}.
It is not difficult to find that
\begin{eqnarray}
q=\partial_2 Q(p,\ y),\ r=\partial_2 R(p,\ z),\  \label{qr}
\end{eqnarray}
with arbitrary functions $Q\equiv Q(p,\ y)$ and $R\equiv R(p,\ z)$ solve the equation \eqref{GHE}, where $\partial_i,\ i=1,\ 2$ are defined as the partial derivatives with respect to the first and second variables, respectively.

Substituting the result \eqref{qr} into the left equations \eqref{GHEpy} and \eqref{GHEpz} yields
\begin{eqnarray}
&&p_y=p_x\partial_{12} Q,\ p_z=p_x\partial_{12} R,\ \partial_{12}\equiv \partial_1\partial_2. \label{pypz}
\end{eqnarray}
It is straightforward to check that the nonlinear system \eqref{pypz} is consistent, i.e., $p_{yz}=p_{zy}$ is identically satisfied.
The general solution of \eqref{pypz} in the implicit form using  hodograph transformation is
\begin{eqnarray}
x+\partial_1 Q+\partial_1 R+T(p,\ t)=0,\  \label{Fp}
\end{eqnarray}
where $T\equiv T(p,\ t)$ is an arbitrary function of $p$ and $t$.
Let $\{p_1,\ q_1,\ r_1\}$ and $\{p_2,\ q_2,\ r_2\}$ be two solutions of \eqref{pypz} possessing the forms
\begin{equation}
x+\partial_1 Q_{i}+\partial_1R_{i}+T_i=0,\  q_i=\partial_2Q_i,\ r_i=\partial_2R_i,\ i=1,\ 2\label{Tpi}
\end{equation}
with six arbitrary functions $Q_i\equiv Q_{i}(p_i,\ y),\ R_i\equiv R_{i}(p_i,\ z)$ and $T_i\equiv T_{i}(p_i,\ t),\ i=1,\ 2.$
One gets new solutions with \eqref{Tpi} via linear superposition only if the balance condition \eqref{GHE12} vanishes. The balance condition \eqref{GHE12} after the substitution of linear superposition formulae \eqref{LS} and \eqref{Tpi} becomes
\begin{eqnarray}
&&a[\partial_{12}(R_2-R_1)][(\partial_2T_1)(\partial_{12}Q_2)
-(\partial_2T_2)(\partial_{12}Q_1)]=\nonumber\\
&&b[\partial_2(T_2-T_1)][(\partial_{12}Q_1)(\partial_{12}R_2)
-(\partial_{12}Q_2)(\partial_{12}R_1)].
\label{PQR}
\end{eqnarray}
Solving out \eqref{PQR} and substituting the result into \eqref{Tpi},
show that linear superposition \eqref{LS} with
\begin{equation}
q_i=\partial_2Q_i=m_i+\beta_y F_i,\ r_i=\partial_2 R_i=n_i+\delta_z F_i,\ i=1,\ 2,
\label{QRT}
\end{equation}
is a nontrivial solution of the GHE,
while $p_i,\ i=1,\ 2$ are implicitly given by
\begin{equation}
x+(\alpha+\beta+\delta) F_{ip_i}+G_i=0,\ i=1,\ 2,\label{Ti}
\end{equation}
where $F_i\equiv F_i(p_i),\ G_i\equiv G_i(p_i),\ m_i\equiv m_i(y),\ n_i\equiv n_i(z),\ \alpha\equiv\alpha(t),\ \beta\equiv\beta(y)$ and $ \delta\equiv\delta(z)$ are arbitrary functions of the indicated variables.

Remarkably enough, the linear superposition solution \eqref{LS} with \eqref{QRT} and \eqref{Ti} should not belong to any solution by redefining the arbitrary functions of $q_1$ (or $q_2$), otherwise, it is referred to as trivial.
For instance, $\{p_i,\ q_i,\ r_i,\ i=1,\ 2\}$ given by
\begin{equation}
p_i=P_i,\ q_i=\int P_{iy}\mbox{\rm dx}+H_i,\ r_i=n_i+\delta_zP_i \label{Pi}
\end{equation}
with $\alpha\equiv \alpha(t),\ \delta\equiv \delta(z),\ P_i\equiv P_i(x+\alpha+\delta,y),\ n_i\equiv n_i(z)$ and $H_i\equiv H_i(y,\ z)$ being arbitrary functions of the indicated variables are two solutions of the GHE \eqref{GHE}. However, the linear superposition solution \eqref{LS} with \eqref{Pi} does not constitute a new solution of \eqref{GHE} because of the arbitrariness of $P_i,\ n_i$ and $F_i$.

Owing to the arbitrariness of $F_i\equiv F_i(p_i)$ and $G_i\equiv G_i(p_i)$, we can not explicitly solve out $p_i$ from the implicit expressions \eqref{Ti}. In other words, we can not prove
\begin{equation}
x+(\alpha+\beta+\delta) F(a_1p_1+a_2p_2)+G(a_1p_1+a_2p_2)=0 \label{FG}
\end{equation}
from \eqref{Ti} for arbitrary $F_i,\ G_i, F$ and $G$.
Thus, the linear superposition solution \eqref{LS} with \eqref{Ti} is actually a wholly new solution of the GHE \eqref{GHE}.

Furthermore, because the Poison bracket \eqref{PB} (and then the GHE \eqref{GHE}) is a bilinear form of the fields $p,\ q$ and $r$, the nonlinear balance condition \eqref{GHE12} should be modified as
\begin{equation}
\sum_{i\neq j}^n\big[a\{r_i,\ p_j\}_{yt}
+b\{r_i,\ q_j\}_{xt}\big]=0\label{GHEn}
\end{equation}
for the linear superposition solution
\begin{equation}
p=\sum_{i=1}^na_ip_i,\ q=\sum_{i=1}^na_iq_i,\ r=\sum_{i=1}^na_ir_i\label{pqrn}
\end{equation}
of the $n$ solutions $\{p_i,\ q_i,\ r_i\}$. It is straightforward to check that the nonlinear balance condition \eqref{GHEn} is indeed satisfied for the following shock wave solutions
\begin{equation}
q_i=\partial_2Q_i=m_i+\beta_y F_i,\ r_i=\partial_2 R_i=n_i+\delta_z F_i,
\label{QRTn}
\end{equation}
and $p_i$ being given by
\begin{equation}
x+(\alpha+\beta+\delta) F_{ip_i}+G_i=0,\ i=1,\ 2,\ \ldots, n, \label{Tn}
\end{equation}
where $F_i\equiv F_i(p_i),\ G_i\equiv G_i(p_i),\ m_i\equiv m_i(y),\ n_i\equiv n_i(z),\ \alpha\equiv\alpha(t),\ \beta\equiv\beta(y)$ and $ \delta\equiv\delta(z)$ are arbitrary functions of the indicated variables.

The method established here will serve as a model in the investigation of linear superposition in more nonlinear systems such as the BKP hierarchy, the dispersionless BKP hierarchy \cite{Hao} and many types of dispersionless Hirota type systems including other types of heavenly equations. We would like to emphasize that what we have obtained is only a kind of linear superposition with special solutions and not the full superposition in the linear theories.

To summarize, analytic difficulties arise because most of the methods we have learned rely on the principle of superposition, a condition that nonlinear systems violate. This letter displays a methodology that extends to nonlinear systems the linear superposition principle which is well developed for linear systems, shows that the implicit shock wave solutions of the general heavenly equation which make the nonlinearities balance, under appropriate conditions obey the superposition principle, and demonstrates the ``impossible" linear superposition principle for nonlinear systems becomes possible. 
In addition to the interesting properties brought by the implicit shock wave solutions of the general heavenly equation, we believe that the linear superposition principle may contribute a totally new insight into the physical nature. Another possible application is nonlinear systems with multi-power nonlinearities where nonlinear cross terms may balance each other allowing for the application of the linear superposition principle on that case as well.

The authors are in debt to the helpful discussions with Professors Hu X. B. and Liu Q. P.. The work was sponsored by the National Natural Science Foundation ofChina (Nos. 11975131 and 10735030) and K. C. Wong Magna Fund in Ningbo University, Zhejiang Provincial Natural Science Foundation of China (No. LQ20A010009) and the General Scientific Research of Zhejiang Province (No. Y201941009).

\nocite{*}

\bibliography{HLS}

\end{document}